\documentclass[aps,showpacs,prl,twocolumn,preprintnumbers,amsmath,amssymb]{revtex4}
\usepackage{graphicx}
\usepackage{dcolumn}
\usepackage{bm}
\def\Bf#1{\mbox{\boldmath{$#1$}}}
\def\bF#1{\mbox{\scriptsize\boldmath{$#1$}}}

\def\Sc#1{\textsc{#1}}

\begin{document}

\preprint{Preprint number: ITP-UU-2004/35}

\title{Comment on ``Quasiparticle Anisotropy and Pseudogap Formation from
the Weak-Coupling Renormalization Group Point of View''}

\author{Behnam Farid}
\email{B.Farid@phys.uu.nl}
\affiliation{Institute for Theoretical
Physics, Department of Physics and Astronomy,
University of Utrecht, \\
Leuvenlaan 4, 3584 CE Utrecht, The Netherlands }

\date{\today}



\maketitle


In their recent Letter, Katanin and Kampf \cite{KK04} (KK)
reported numerical results for the self-energy $\Sigma({\Bf
k};\varepsilon)$, $\varepsilon \in \mathbb{R}$, of the single-band
Hubbard Hamiltonian (hereafter `single-band' will be implicit) in
two space dimensions (i.e. $d=2$), obtained through employing the
functional renormalization-group (fRG) formalism (see references
in \cite{KK04}, and \cite{KB01}) at the one-loop level. Several of
the results by KK are in full conformity with the {\em exact}
formal results reported earlier in \cite{BF03,BF04a,BF04b}. This,
as we shall elaborate below, strengthens one's confidence in the
reliability of the fRG in dealing with models of
strongly-correlated fermions.

\begin{figure}[t!]
\includegraphics[width=2.0in, angle=-90]{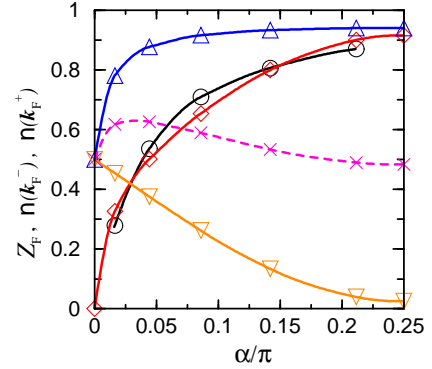}
\vspace{-0.3cm} \caption{\label{f1} $Z_{\Sc f}$ as calculated by
KK ($\bigcirc$) and according to the exact expression in the 
text with $b^-=0.0912$, $b^+ = -1.4158$ ($\Diamond$), 
corresponding to $U/t=2$ and the vH filling associated with 
$t'/t=0.1$. $\alpha \equiv \widehat{{\Bf k}_{\Sc f}, (\pi,0)}$. 
The apparent deviation between the two results reflects the 
isotropy of $b^{\mp}$ {\em assumed} here. Using the above 
$b^{\mp}$, we also present ${\sf n}({\Bf k}_{\Sc f}^-)$
($\bigtriangleup$) and ${\sf n}({\Bf k}_{\Sc f}^+)$
($\bigtriangledown$) obtained from the exact expressions in the 
text, as well as $[{\sf n}({\Bf k}_{\Sc f}^-) + {\sf n}({\Bf k}_{\Sc
f}^+)]/2$ ($\times$) whose deviation from $1/2$ in the present
case is indicative of the underlying metallic state {\em not}
being a Fermi liquid \protect\cite{BF03} (leaving aside $Z_{\Sc
f}=0$ at $\alpha=0$). \vspace{-0.2cm}}
\end{figure}

{\it (i) The quasi-particle (qp) weight $Z_{\Sc f}$ at the Fermi
surface.} --- Amongst other things, KK showed that \cite{KK04}
``At van Hove (vH) band fillings and at low temperatures, the
quasiparticle weight along the Fermi surface (FS) continuously
vanishes on approaching the $(\pi,0)$ point \dots.'' In
\cite{BF03} we have obtained a general expression for the
momentum-distribution function ${\sf n}({\Bf k})$ at ${\Bf k}={\Bf
k}_{\Sc f}^{\mp}$, i.e. infinitesimally in- and outside Fermi
sea (here we suppress the spin indices encountered in 
\cite{BF03,BF04a,BF04b}), which on the basis of the available 
numerical results at the time we have shown to reduce to (Eq.~(99) 
in \cite{BF03}): ${\sf n}({\Bf k}_{\Sc f}^{\mp}) = (a + 
U b^{\mp})/(a + 2 U b^{\mp})$, where $a$ and $b^{\mp}$ are components 
of vectors ${\Bf a}({\Bf k}_{\Sc f}) \equiv {\Bf\nabla}_{\bF k}
\varepsilon_{\bF k}\vert_{{\bF k} = {\bF k}_{\Sc f}}$ and 
${\Bf b}({\Bf k}_{\Sc f}^{\mp})$ along the outward unit vector 
normal to Fermi surface ${\cal S}_{\Sc f}$ at ${\Bf k}={\Bf
k}_{\Sc f}$; here $\varepsilon_{\bF k}$ is the non-interacting
energy dispersion and ${\Bf b}({\Bf k})$ the gradient of a
well-defined scalar ground-state (GS) correlation function.
Stability of the latter GS is tantamount to the satisfaction of
\cite{BF03} $b^- > a/(U \Lambda^-)$, $b^+ < -a/U$, where
$\Lambda^- \equiv {\sf n}({\Bf k}_{\Sc f}^-)/[1 -{\sf n}({\Bf
k}_{\Sc f}^-)] > 0$. Our above expression for ${\sf n}({\Bf
k}_{\Sc f}^{\mp})$ makes explicit that (a) ${\sf n}({\Bf k}_{\Sc
f}^{\mp}) \gtreqqless 1/2$, (b) ${\sf n}({\Bf k}_{\Sc f}^{\mp})
\to 1/2$, i.e. $Z_{\Sc f} \equiv {\sf n}({\Bf k}_{\Sc f}^-)
- {\sf n}({\Bf k}_{\Sc f}^+) \to 0$, for $U b^{\mp} \to \infty$, and
(c) ${\sf n}({\Bf k}_{\Sc f}^{\mp}) = 1/2$, i.e. $Z_{\Sc f} = 0$,
for $a=0$, that is for ${\Bf k}_{\Sc f}$ at vH points. In
Fig.~\ref{f1} we compare our results for $Z_{\Sc f}$ with those
determined by KK \cite{KK04}, {\em disregarding} the dependence of
$b^{\mp}$ on the direction of ${\Bf k}_{\Sc f}$.

{\it (ii) The single-particle spectral function $A({\Bf
k};\varepsilon)$ for ${\Bf k}$ in the pseudogap (PG) region.} ---
KK observed that \cite{KK04} ``The qp weight suppression [for
${\Bf k}_{\Sc f} \to (\pi,0)$] is accompanied by the growth of two
additional incoherent peaks in the spectral function, from which
an anisotropic pseudogap originates.'' On general theoretical
grounds, in \cite{BF03, BF04a} we have shown that the {\em
singular} nature of ${\sf n}({\Bf k})$ at {\em all} ${\Bf k} \in 
{\cal S}_{\Sc f}^{(0)}$ (see {\it (iii)} below) implies that (see
Sec.~10 in \cite{BF03}) $Z_{\Sc f} \to 0$, for ${\Bf k} \to$ PG,
{\em must} necessarily be accompanied by at least two {\em
resonant} peaks (to be distinguished from qp peaks) in $A({\Bf
k};\varepsilon)$, one {\em strictly below} and one {\em strictly
above} the Fermi energy $\varepsilon_{\Sc f}$. Here ${\cal S}_{\Sc
f}^{(0)}$ is the Fermi surface associated with $\varepsilon_{\bF
k}$ (see {\it (iii)} below).

{\it (iii) Fermi surface non-deformation.} --- For models
involving solely contact-type interaction, we have shown that
\cite{BF03} ${\cal S}_{\Sc f} \subseteq {\cal S}_{\Sc f}^{(0)}$;
PG consists of those points of ${\cal S}_{\Sc f}^{(0)}$, if any,
which do not belong to ${\cal S}_{\Sc f}$ \cite{BF03,BF04a}.
Interestingly, ${\cal S}_{\Sc f} \subseteq {\cal S}_{\Sc f}^{(0)}$
turns out to be the working hypothesis for many calculations,
amongst which those by KK \cite{KK04}.

We should like to emphasize that the above-mentioned results,
cited from \cite{BF03,BF04a,BF04b}, are {\em not} restricted to
the weak-coupling limit; the only constraint for the validity of
these results is the uniformity of the underlying GSs. We further
point out that, {\em in principle}, depending on the values of
$d$, $U/t$, $t'/t$, etc., some other `universality classes' for
the uniform metallic GSs of the Hubbard Hamiltonian (explicitly
considered in \cite{BF03,BF04a}) may/do become viable than that 
dealt with in this Comment.

I thank Professors Henk Stoof and Peter Kopietz for discussions.

\bibliographystyle{apsrev}


\end{document}